\def\calb  {{\cal B}}
\def\calc  {{\cal C}}
\def\cald  {{\cal D}}
\def\calf  {{\cal F}}
\def\calh  {{\cal H}}
\def\call  {{\cal L}}
\def\dl            {\mathbb }
\newcommand\erf[1] {(\ref{#1})}
\newcommand\hsp[1] {\mbox{\hspace{#1 em}}}
\def\ii            {{\rm i}}
\newcommand\nxt[1] {\\\raisebox{.12em}{\rule{.35em}{.35em}}\hsp{.6}#1}
\def\rmd           {{\rm d}}
\def\zet           {{\dl Z}}

\documentclass[12pt]{article} \usepackage{amssymb,amsfonts,latexsym}
\begin{document}

%%%%%%%%%%%%%%%%%%%%%%%%%%%%%%%%%%%%%%%%%%%%%%%%%%%%%%%%%%%%%%%%%%%%%%%
\begin{flushright}  {~} \\[-1cm]
{\sf hep-th/0012045}\\{\sf PAR-LPTHE 00-45}\\[1mm]
{\sf December 2000} \end{flushright}

\begin{center} \vskip 14mm
{\Large\bf D-BRANES IN GROUP MANIFOLDS AND} \\[4mm]
{\Large\bf FLUX STABILIZATION}\\[20mm]
{\large Christoph Schweigert}
\\[8mm]
LPTHE, Universit\'e Paris VI~~~{}\\
4 place Jussieu\\ F\,--\,75\,252\, Paris\, Cedex 05
\end{center}
\vskip 18mm
\begin{quote}{\bf Abstract}\\[1mm]
We consider D-branes in group manifolds, from the point of view
of open strings and using the Born-Infeld action on the brane
worldvolume. D-branes correspond to certain integral (twined) conjugacy 
classes. We explain the integrality condition on the conjugacy classes
in both approaches. In the Born-Infeld description, the D-brane worldvolume
is stabilized against shrinking by a subtle interplay of quantized U(1) fluxes
and the non-triviality of the B-field.
\end{quote}

\bigskip

\section{D-branes}

The role of D-branes in the description of solitonic sectors of string
theories is by now well established. Much insight has been gained from
the fact that D-branes have two complementary descriptions:
\nxt on the one hand, they correspond to conformally invariant
     boundary conditions of open strings
\nxt on the other hand, as solitonic objects, they can be described
     by a worldvolume action of Born-Infeld form.

Here, we are interested in D-branes in backgrounds with non-vanishing
values for the metric $G$ and the Kalb-Ramond field $B$ of
target space. In particular, we consider a case where $B$ is not
even a closed two-form. We will see that the non-triviality of $B$ 
requires a {\em quantization} of the possible positions of the D-brane.

Our test case are strings on group manifolds, so-called WZW models;
for these backgrounds exact results using two-dimensional conformal
field theory are available. Our goal is to see to what extent these results 
can be derived from {\em classical} geometry. 

For simplicity, we mostly restrict ourselves to the case of $G=SU(2)$. 
Topologically, $SU(2)$ is a three-sphere $S^3$. We write it as a union of 
two-spheres with coordinates $0\leq\theta\leq\pi$, $0\leq\phi\leq2\pi$ that are 
parametrized by an angle $0<\psi<\pi$, with the two elements $+\bf 1$ and 
$-\bf 1$ of $SU(2)$ added as the ``north pole'' and the ``south pole'' of 
$S^3$. In these coordinates, the metric reads
$$ \rmd s^2 = k\alpha^\prime \left[ d\psi^2 + \sin^2\psi \Bigl( d\theta^2 +
{\rm sin}^2\theta\;  d\phi^2 \Bigr)\right]\ , $$
and the Neveu-Schwarz three-form  background field is 
$$ H  \equiv {dB} = {2k\alpha^\prime}\; \sin^2\psi\;
{\rm sin}\theta\; d\psi\; d\theta\; d\phi\ . $$
The corresponding two-form potential has a Dirac string singularity,
which we choose at $\psi=\pi$:
\begin{equation} B =  k\alpha^\prime
\left( \psi - {\sin2\psi\over 2}\right)\;
\sin\theta\; d\theta\; d\phi \ . \label B \end{equation}
Notice that the Dirac string breaks translation invariance on the group
manifold. The two-spheres of ``constant latitude'' can also be characterized as
conjugacy classes: they consist of all elements of $SU(2)$ that are
of the form 
$$ \calc (h) = \{ \, g h g^{-1} \,|\, g\in SU(2) \,\} $$
for some fixed diagonal matrix $h$.

\section{Open string analysis}

The background we have just described possesses a non-abelian current algebra
as its symmetry so that algebraic methods allow to obtain exact results 
\cite{fffs}. We wish to describe those
conformally invariant boundary conditions for which left moving and right
moving currents are connected at the boundary of the world sheet by the
action of an automorphism $\omega$ of $G$. This implies that the corresponding
boundary state $|\omega,\alpha\rangle$ is constructed from twisted 
Ishibashi states $|\lambda,\omega\rangle\rangle\in\calh_\lambda\otimes
\calh_{\lambda^+}$ that obey
$$ [J^a_n \otimes 1 + 1\otimes \omega(J^a_{-n}) ] \,
|\lambda\omega\rangle\rangle = 0 \, . $$

The boundary state has been computed in \cite{bifs} and reads,
in a suitable normalization of the bulk fields:
$$ |\omega,\alpha\rangle = \sum \, \chi_\lambda^\omega(h_\alpha) \,
|\lambda\omega\rangle\rangle $$
where $\chi_\lambda^\omega$ is the so-called twining character
\cite{fusS3} and $h_\alpha = \exp(2\pi\ii y_\alpha)$. 
$y_\alpha$ takes its values in a finite set of symmetric weights.

This singles out a finite set of D-branes; their geometry can be
directly tested using bulk fields. Consider first the case of flat
backgrounds: from the one-point functions on a disc with boundary
condition $\beta$:
\begin{equation} 
G^{ij} (\vec q) = \langle \beta | \alpha_{-1}^i\otimes \alpha_{-1}^j|\vec 
q\rangle \label1 \end{equation}
one obtains \cite{dfpslr}, after a Fourier transform, the fluctuations
of the metric, dilaton and Kalb-Ramond field in the classical D-brane
solution.

The symmetries of a free background form an abelian current algebra.
The generalization of \erf1 to the non-abelian case reads
$$G^{ab}_{\omega,\alpha} (v\otimes \tilde v) = 
\langle \omega,\alpha| J^a_{-1}\otimes J^b_{-1} | v\otimes \tilde v\rangle
$$ 
where $v\otimes\tilde v$ is a vector in
$ \calf_k = \oplus_{\lambda\in P_k} \, \bar\calh_\lambda\otimes
\bar\calh_{\lambda^+} \,; $
the sum is over integrable highest weights at level $k$. We identify
$\calf_k$ with a subspace of the space 
$ \calf = \oplus_{\lambda} \, \bar\calh_\lambda\otimes \bar\calh_{\lambda^+}$
of functions on $G$. Then $G^{ab}_{\omega,\alpha}$ can be interpreted 
\cite{fffs} as a distribution on $G$:
$$ G^{ab}_{\omega,\alpha}(g) = - \kappa^{ab} \sum_{\lambda\in P^\omega_k} 
\chi_\lambda^\omega(h_\alpha)^* \chi_\lambda^\omega(g) + \ldots $$
where we have dropped antisymmetric terms. Thus the fluctuations of the
metric are proportional to the Killing form $\kappa^{ab}$ and
concentrated at the so-called {\em twined conjugacy class}
$$ \calc^\omega(h) = \{ \, g h \,\omega(g)^{-1}\,  |\,  g\in G \, \} \,. $$
For a more detailed description of the geometry of these subspaces we
refer to \cite{fffs}; from now on we restrict ourselves to the
case of trivial automorphism $\omega=1$.

\section{Space-time analysis}

We now consider the (bosonic) Born-Infeld action
\begin{equation}
\int_{\calb} \rmd^p \xi \sqrt{\det(G + B + 2\pi\alpha' F)} 
\label{dbi} \end{equation}
on the worldvolume $\calb$ of the brane. Its form generalizes the action for 
minimal surfaces: it depends also on the antisymmetric tensor field $B$. 
Moreover, one has to choose a connection $A$ with field strength $F$ on $\calb$.
We {\em assume} that classical geometry captures the essential features of the 
problem; as a consequence the flux $\int_{\calb} F$ is quantized.

This point deserves a careful discussion, since one might have tried
to impose a quantization of the gauge invariant field strength
$\calf := B + 2\pi\alpha' F$. A first simple remark is that under the
gauge transformations
$$ \delta B = 2\pi\alpha'\rmd\Lambda \qquad \delta A = -\Lambda $$
integrality of the flux of $F$ is preserved, although its integral value
can be changed by large gauge transformations. 

One can also see the quantization of $F$ directly from the following 
worldsheet argument \cite{pelc}. Apart from the kinetic term, the WZW action 
on a worldsheet $\Sigma$ contains a bulk term with the pullback of $B$
and a boundary term with the pull-back of the gauge field $A$ on the brane:
\begin{equation}
\call' = \int_\Sigma B - 2\pi \alpha \int_{\partial\Sigma} A \, . 
\label{action} \end{equation}
Let us suppose for simplicity that $\Sigma$ has a single boundary component.
We can close $\Sigma$ by choosing a disc $\cald$ in the world volume of
the brane. We find
$$ 
\call' = \int_{\Sigma\cup \cald} B - \int_{\cald} B - 2\pi\alpha' 
\int_\cald F 
= \int_{\calb} H - \int_\cald \calf  \, , $$
where we have chosen a three-manifold $\calb$ with boundary $\Sigma\cup \cald$.
Our considerations should be independent from the choices of $\cald$ and
$\calb$. Changing $\calb$, but keeping $\cald$ fixed, leads, as usual, to
the requirement that the level $k$ should be an integer. On the other hand,
changing $\cald$ to $\cald'$ requires to change $\calb$ to $\calb' =
\calb\cup\tilde\calb$, where $\tilde\calb$ is a full ball, bounded by
$\cald\cup(-\cald')$. The difference of the action \erf{action} then reads
$$ 
\int_{\tilde\calb} H - \int_{\partial\tilde\calb} 2\pi\alpha' F + B 
= - 2\pi\alpha' \int_{\partial\tilde\calb} F \,, $$
which, taking into account the correct normalization of the metric, leads
to the quantization of $F$, rather than to a quantization of $\calf$. 
A similar argument has been given in \cite{alsc2}, although in a 
different order: while we first fix the correct topology and then impose
the equations of motion for the Born-Infeld theory, the authors of \cite{alsc2}
first imposed (a stronger requirement implying) conformal invariance 
on the boundary, i.e.\ the equations of motion of the string-worldsheet 
theory, to obtain conjugacy classes and then used the topological constraint to 
find integral conjugacy classes.

To find extrema of the action \erf{dbi}, we use a ``mini-superspace''
approach: for $\calb$, we consider only submanifolds of the form 
$\psi={\rm const}$ and we fix the connection to 
$$ F = \rmd A = -\frac n2 \sin\theta \rmd\theta\wedge \rm\phi \,, $$
where $n\in\zet$ gives the flux of $F$.
As a function of the single variable $\psi$, the energy then reads
$$ E_n(\psi) = 4\pi k\alpha' (\sin^4\psi + (\psi-\frac{\sin2\psi}2 - 
\frac{\pi n}k)^2)^{1/2} \, , $$
which has a minimum for $\psi_n= \pi n/ k$. At this minimum, the mass
is $M_n=4\pi k\alpha'T_{(2)}\sin\frac{\pi n}k$ which coincides {\em exactly}
with the CFT result. This is truly remarkable, since the theory in question
is {\em not} supersymmetric. We see that the quantization
of the possible flux $n$ over $S^2$ implies a quantization of the position
$\psi$ of the D-brane. The only term in $E$ that is not a trigonometric 
function of $\psi$ comes from the $B$-field \erf B; this is the way the
non-triviality of $B$ enters in the determination of the position of the
brane and conspires with the flux on the D-brane worldvolume to stabilize
the D-brane at a finite size.

The charge with respect to the gauge invariant field strength $\calf$ turns
out to be 
$$ Q_n = T_{(2)} \int_{S^2} B + 2\pi F = 2\pi k \alpha' T_{(2)}
\sin\frac{2\pi n}k \, , $$
again in {\em exact} agreement with the CFT result.
Note that these charges are {\em not} rationally related. Notice, though,
that in the limit of large level $k$ (and thus of week curvature)
$Q_n$ approaches the charge of $n$ free D-particles. It is therefore
tempting to interpret a D2-brane at $\psi_n= \pi n/k$ as a bound state of 
$n$ D0-branes. This idea has recently received some attention. To turn it into
a quantitative argument, a careful understanding of the non-abelian Born-Infeld
action \cite{myer} is needed, though.

One can also compute the spectrum of the quadratic fluctuations. For 
$j=0,1,\ldots$, one finds states of
$$ m^2 = \frac{j(j+1)}{k\alpha'} $$
in representations of spin $(j-1)\oplus(j+1)$. In particular, one finds
a triplet of zero-modes corresponding exactly to the three rotations in $SU(2)$.
All other values of $m^2$ are positive, which confirms the stability of our 
solution.

A comparison with the CFT results shows that we get the right number of
branes at the right locations and with the correct energy. The charge $Q_n$,
moreover, reproduces the correct Ramond-Ramond charge in a supersymmetrized
WZW model.  There is only a discrepancy for the spectrum of quadratic 
fluctuations: they should correspond to the open string states which, according
to the exact CFT result, come in the affine representation
$ \oplus_{j=0}^{[k/2]} \, \calh_j \,. $
In both approaches only states with integral spin appear. The Born-Infeld
action only sees certain affine descendants. Moreover, the cut-off
by the level $k$ is not found in the Born-Infeld approach. One possible
explanation is that high spin states, like high momentum states in a free
theory, test the ultraviolet structure of space-time. In this space-time
analysis, we work with a classical, smooth space-time. It would be, however, 
interesting to see whether higher order fluctuations could lead to a 
level-dependent truncation.

\section{Conclusions}

It is quite remarkable (and still not really understood) why the Born-Infeld 
action gives essentially the {\em exact} CFT results, in spite of the fact 
that the theory is not
supersymmetric. In the case of $G=SU(2)$, one might be tempted to find an
explanation by embedding the SU(2)-theory into the background describing
$k$ NS5-branes. 

However, exactness of the results should generalize to arbitrary simple 
compact Lie groups. A first check is provided by the following result: the 
number of coordinates needed to fix the position of a brane with $\omega=1$ has
to be equal to the number of independent U(1)-fluxes on the brane worldvolume
$G/T$, where $T$ is a maximal torus of $G$. This holds indeed true: the fluxes 
take their values in the lattice $H^2(G/T,\zet)$, whose rank equals the rank 
of $G$, which is the number of coordinates needed to fix the brane world 
volume. One can go even further: according to the algebraic theory, possible 
D-branes correspond to integrable highest weights of $G$. Indeed, 
Borel-Bott-Weil 
theory allows to associate to each integral weight of $G$ a unique line bundle 
over $G/T$ which should play the role of the gauge bundle on the brane.

Other open questions concern more general conformally invariant boundary 
conditions. From algebraic investigations, it is known that a WZW theory 
possesses many more conformally invariant boundary conditions, where, 
however, left movers
and right movers are not any longer connected by an automorphism $\omega$.
On the geometric side, one should also consider branes with more general
topology, e.g.\ for $G=SU(2)$ D2-branes of higher genus. Both approaches and
their relation remain to be explored.

\section*{Acknowledgments}
I would like to thank the organizers of the IXth Marcel Grossmann
meeting for their invitation to present these results and
C.\ Bachas, L.\ Birke, M.\ Douglas, G.\ Felder, J.\ Fr\"ohlich
and J.\ Fuchs for enjoyable collaborations.

\end{document}